%% file: main.tex
\documentclass[sigconf]{acmart} 

\usepackage{amsmath}
\usepackage[ruled,vlined,linesnumbered]{algorithm2e}
\usepackage{multirow}
\usepackage{subcaption}
\usepackage{kotex}
\usepackage{mathtools}
\DeclarePairedDelimiter{\abs}{\lvert}{\rvert}
\usepackage{enumitem}
\setlist[itemize]{leftmargin=3mm}
\usepackage[skip=2pt]{caption} 
\usepackage{stackengine}

\AtBeginDocument{%
  \providecommand\BibTeX{{%
    \normalfont B\kern-0.5em{\scshape i\kern-0.25em b}\kern-0.8em\TeX}}}

\setcopyright{iw3c2w3}

\begin{document}

\copyrightyear{2021}
\acmYear{2021} 
\acmConference[WWW '21]{Proceedings of the Web Conference 2021}{April 19--23, 2021}{Ljubljana, Slovenia} 
\acmBooktitle{Proceedings of the Web Conference 2021 (WWW '21), April 19--23, 2021, Ljubljana, Slovenia}
\acmPrice{}
\acmDOI{10.1145/3442381.3449878}
\acmISBN{978-1-4503-8312-7/21/04}
\settopmatter{printacmref=true}

\title{Bidirectional Distillation for Top-$K$ Recommender System}

\author{Wonbin Kweon, SeongKu Kang, Hwanjo Yu\*}
\affiliation{%
   \institution{Pohang University of Science and Technology, South Korea}
   \country{}
   \{kwb4453, seongku, hwanjoyu\}@postech.ac.kr
}
\authornotemark[0]
\authornote{Corresponding Author}


\begin{abstract}
\input{0abst.tex}
\end{abstract}



\maketitle

\section{Introduction}
\input{1intro.tex}

\section{Analysis: Teacher vs Student}
\input{2Analysis}

\section{Problem Formulation}
\input{3pre}

\section{Method}
\input{4model.tex}

\section{Experiments}
\input{5ex.tex}

\section{Related Work}
\input{6rw}

\section{Conclusion}
\input{7con.tex}

\section*{Acknowledgements}
This work was supported by the NRF grant funded by the MSIT (South Korea, No. 2020R1A2B5B03097210), and the IITP grant funded by the MSIT (South Korea, No. 2018-0-00584, 2019-0-01906).

\bibliographystyle{ACM-Reference-Format}
\bibliography{main}

\pagebreak
\newpage
\appendix
\input{8app.tex}

\end{document}

%% file: 0abst.tex
Recommender systems (RS) have started to employ knowledge distillation, which is a model compression technique training a compact model (student) with the knowledge transferred from a cumbersome model (teacher).
The state-of-the-art methods rely on unidirectional distillation transferring the knowledge only from the teacher to the student, with an underlying assumption that the teacher is always superior to the student.
However, we demonstrate that the student performs better than the teacher on a significant proportion of the test set, especially for RS.
Based on this observation, we propose \textit{Bidirectional Distillation} (BD) framework whereby both the teacher and the student collaboratively improve with each other. 
Specifically, each model is trained with the distillation loss that makes to follow the other’s prediction along with its original loss function.
For effective bidirectional distillation, we propose \textit{rank discrepancy-aware sampling} scheme to distill only the informative knowledge that can fully enhance each other.
The proposed scheme is designed to effectively cope with a large performance gap between the teacher and the student.
Trained in the bidirectional way, it turns out that both the teacher and the student are significantly improved compared to when being trained separately.
Our extensive experiments on real-world datasets show that our proposed framework consistently outperforms the state-of-the-art competitors.
We also provide analyses for an in-depth understanding of BD and ablation studies to verify the effectiveness of each proposed component.

%% file: 1intro.tex
Nowadays, the size of recommender systems (RS) is continuously increasing, as they have adopted deep and sophisticated model architectures to better understand the complex relationships between users and items \cite{rd18, cd19}.
A large recommender with many learning parameters usually has better performance due to its high capacity, but it also has high computational costs and long inference time.  
This problem is exacerbated for web-scale applications having numerous users and items, since the number of the learning parameters increases proportionally to the number of users and items.
Therefore, it is challenging to adopt such a large recommender for real-time and web-scale platforms. 

To tackle this problem, a few recent work \cite{rd18, cd19, DERRD} has adopted \textit{Knowledge Distillation} (KD) to RS.
KD is a model-agnostic strategy that trains a compact model (student) with the guidance of a pre-trained cumbersome model (teacher).
The distillation is conducted in two stages;
First, the large teacher recommender is trained with user-item interactions in the training set with binary labels (i.e., $0$ for unobserved interaction and $1$ for observed interaction.).
Second, the compact student recommender is trained with the recommendation list predicted by the teacher along with the binary training set. 
Specifically, in \cite{rd18}, the student is trained to give high scores on the top-ranked items of the teacher's recommendation list. 
Similarly, in \cite{cd19}, the student is trained to imitate the teacher's prediction scores with particular emphasis on the high-ranked items in the teacher's recommendation list.
The teacher's predictions provide additional supervision, which is not explicitly revealed from the binary training set, to the student.
By distilling the teacher's knowledge, the student can achieve comparable performance to the teacher with fewer learning parameters and lower inference latency.

\begin{figure*}[t]
\begin{subfigure}[t]{0.495\linewidth}
    \includegraphics[width=0.5\linewidth]{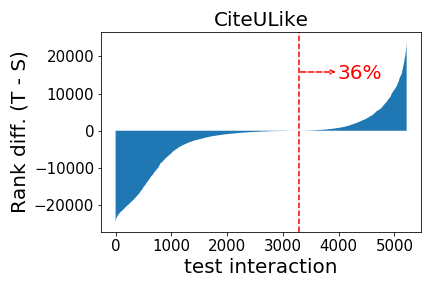}
    \includegraphics[width=0.5\linewidth]{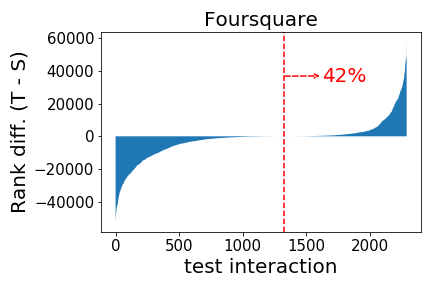}
    \caption{Rank difference between teacher and student}
\end{subfigure} 
\begin{subfigure}[t]{0.495\linewidth}
    \includegraphics[width=0.5\linewidth]{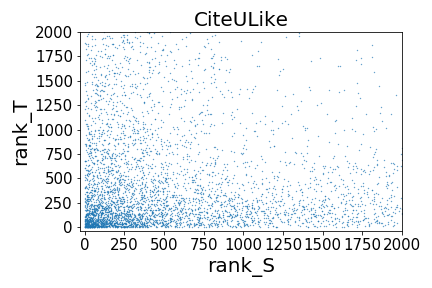}
    \includegraphics[width=0.5\linewidth]{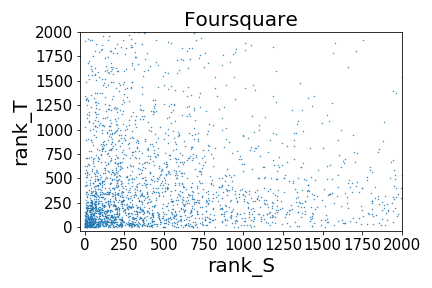}
    \caption{Rank comparison between teacher and student}
\end{subfigure}
\caption{Performance analyses of the teacher and the student. In (a), rank differences are computed on the test interactions and the test interactions (x-axis) are sorted in increasing order by the rank difference. In (b), each dot represents a sampled unobserved interaction.}
\end{figure*}

Despite their effectiveness, they still have some limitations.
First, they rely on \textit{unidirectional} knowledge transfer, which distills knowledge only from the teacher to the student, with an underlying assumption that the teacher is always superior to the student.
However, based on the in-depth analyses provided in Section 2, we argue that the knowledge of the student also could be useful for the teacher.
Indeed, we demonstrate that the teacher is not always superior to the student and further the student performs better than the teacher on a significant proportion of the test set.
In specific, the student recommender better predicts 36$\sim$42\% of the test interactions than the teacher recommender.
These proportions are remarkably large compared to 4$\sim$9\% from our experiment on the computer vision task.
In this regard, we claim that both the student and the teacher can take advantage of each other's complementary knowledge, and be improved further.
Second, the existing methods have focused on distilling knowledge of items ranked highly by the teacher.
However, we observe that most of the items ranked highly by the teacher are already ranked highly by the student (See Section 2 for details).
Therefore, merely high-ranked items may not be informative to fully enhance the other model, which leads to limiting the effectiveness of the distillation.

We propose a novel \textbf{\underline{B}}idirectional \textbf{\underline{D}}istillation (BD) framework for RS.
Unlike the existing methods that transfer the knowledge only from the pre-trained teacher recommender, both the teacher and the student transfer their knowledge to each other within our proposed framework.
Specifically, they are trained simultaneously with the distillation loss that makes to follow the other’s predictions along with the original loss function.
Trained in the bidirectional way, it turns out that both the teacher and the student are significantly improved compared to when being trained separately.
In addition, the student recommender trained with BD achieves superior performance compared to the student trained with conventional distillation from a pre-trained teacher recommender.

For effective bidirectional distillation, the remaining challenge is to design an \textit{informative} distillation strategy that can fully enhance each other, considering the different capacities of the teacher and the student.
As mentioned earlier, items merely ranked highly by the teacher cannot give much information to the student.
Also, it is obvious that the student's knowledge is not always helpful to improve the teacher, as the teacher has a much better overall performance than the student.
To tackle this challenge, we propose \textit{rank discrepancy-aware sampling} scheme differently tailored for the student and the teacher.
In the scheme, the probability of sampling an item is defined based on its rank discrepancy between the two recommenders.
Specifically, each recommender focuses on learning the knowledge of the items ranked highly by the other recommender but ranked lowly by itself.
Taking into account the performance gap between the student and the teacher, we enable the teacher to focus on the items that the student has very high confidence in, whereas making the student learn the teacher's broad knowledge on more diverse items.

The proposed BD framework can be applicable in many application scenarios.
Specifically, it can be used to maximize the performance of the existing recommender in the scenario where there is no constraint on the model size (in terms of the teacher), or to train a small but powerful recommender as targeted in the conventional KD methods (in terms of the student).
The key contributions of our work are as follows:
\begin{itemize}[leftmargin=*]
    \item Through our exhaustive analyses on real-world datasets, we demonstrate that the knowledge of the student recommender also could be useful for the teacher recommender.
    Based on the results, we also point out the limitation of the existing distillation methods for RS under the assumption that the teacher is always superior to the student.
    \item We propose a novel bidirectional KD framework for RS, named BD, enabling that the teacher and the student can collaboratively improve with each other during the training.
    BD also adopts the rank discrepancy-aware sampling scheme differently designed for the teacher and the student considering their capacity gap.
    \item We validate the superiority of the proposed framework by extensive experiments on real-world datasets. 
    BD considerably outperforms the state-of-the-art KD competitors.
    We also provide both qualitative and quantitative analyses to verify the effectiveness of each proposed component\footnote{We provide the source code of BD at \url{https://github.com/WonbinKweon/BD_WWW2021}}.
\end{itemize}

%% file: 2Analysis.tex
In this section, we provide detailed analyses of comparison between the teacher and the student on two real-world datasets: CiteULike \cite{CUL13} and Foursquare \cite{FS14}.
Following \cite{cd19}, we hold out the last interaction of each user as test interaction, and the rest of the user-item interactions are used for training data
(see Section 5.1 for details of the experiment setup).
We use NeuMF \cite{ncf17} as a base model, and adopt NeuMF-50 as the teacher and NeuMF-5 as the student. The number indicates the dimension of the user and item embedding; 
the number of learning parameters of the teacher is 10 times bigger than that of the student, and accordingly, the teacher shows a much better overall performance than the student (reported in Table 2).
Note that the teacher and the student are trained separately without any distillation technique in the analyses.

\vspace{3pt} \noindent \textbf{The teacher is not always superior to the student.}
Figure 1a shows bar graphs of the rank difference between the teacher and the student on the test interactions (the bars in the graphs are sorted in increasing order along with the x-axis).
The rank difference on a test interaction $(u, i)$ is defined as follows:
\begin{equation}
\text{Rank diff.}^{u}(i) = rank_T^{u}(i) - rank_S^{u}(i),
\end{equation}
where $rank_T^{u}(i)$ and $rank_S^{u}(i)$ denote the ranks assigned by the teacher and the student on the item $i$ for the user $u$ respectively, and $rank_{*}^{u}(i)=1$ is the highest ranking.
If the rank difference is bigger than zero, it means that the student assigns a higher rank than the teacher on the test interaction; the student better predicts the test interaction than the teacher.
We observe that the student model assigns higher ranks than the teacher model on the significant proportion of the entire test interactions: 36\% for CiteULike and 42\% for Foursquare.
These proportions are remarkably large compared to 4\% for CIFAR-10 and 9\% for CIFAR-100 from our experiments on computer vision task\footnote{We compare the predictions of the teacher and the student on the test set. We adopt ResNet-80 as the teacher and ResNet-8 as the student \cite{resnet16} and train them separately for the image classification task on CIFAR-10 and CIFAR-100 dataset \cite{cifar10}.}.

The results raise a few questions.
Why the student can perform better than the teacher on the test data and why this phenomenon is intensified for RS?
We provide some possible reasons as follows:
Firstly, not all user-item interactions require sophisticated calculations in high-dimensional space for correct prediction.
As shown in the computer vision \cite{overthinking19}, a simple image without complex background or patterns can be better predicted at the lower layer than the final layer in a deep neural network.
Likewise, some user-item relationships can be better captured based on simple operations without expensive computations.
Moreover, unlike the image classification task, RS has very high ambiguity in nature;
in many applications, the user’s feedback on an item is given in the binary form: 1 for observed interaction, 0 for unobserved interaction.
However, the binary labels do not explicitly show the user's preference.
For example, “0” does not necessarily mean the user’s negative preference for the item.
It can be that the user may not be aware of the item.
Due to the ambiguity in the ground-truth labels, the lower training loss does not necessarily guarantee a better ranking performance.
Specifically, in the case of NeuMF, which is trained with the binary cross-entropy loss,
the teacher can better minimize the training loss and thus better discriminate the observed/unobserved interactions in the training set. 
However, this may not necessarily result in better predicting user's preference on all the unobserved items due to the ambiguity of supervision.

\vspace{3pt} \noindent \textbf{The high-ranked items are not that informative as expected.}
Figure 1b shows the rank comparison between the teacher and the student.
We plot the high-ranked items in the recommendation list from the teacher and that from the student for all users\footnote{We sample the items by using the rank-aware sampling scheme suggested in \cite{cd19}. Note that in \cite{cd19}, the sampled items are used for the distillation.}.
Each point corresponds to $(rank^u_S(i), rank^u_T(i))$.
We observe that most of the points are located near the lower-left corner, which means that most of the items ranked highly by the teacher are already ranked highly by the student.
In this regard, unlike the motivation of the previous work \cite{rd18, cd19}, items merely ranked highly by the teacher may be not informative enough to fully enhance the student, and thus may limit the effectiveness of the distillation.
We argue that each model can be further improved by focusing on learning the knowledge of items ranked highly by the other recommender but ranked lowly by itself (e.g., points located along the x-axis and y-axis).

In summary, the teacher is not always superior to the student, so the teacher can also learn from the student.
Their complementarity should be importantly considered for effective distillation in RS.
Also, the distillation should be conducted with consideration of the rank discrepancy between the two recommenders.
Based on the observations, we are motivated to design a KD framework that the teacher and the student can collaboratively improve with each other based on the rank discrepancy.

%% file: 3pre.tex
Let the set of users and items be denoted as $\mathcal{U}=\{u_1, u_2,...,u_n\}$ and $\mathcal{I}=\{ i_1, i_2,...,i_m \}$, where $n$ is the number of users and $m$ is the number of items.
Let $\textbf{R} \in \{0,1\}^{n \times m}$ be the user-item interaction matrix, 
where $r_{ui}=1$ if the user $u$ has an interaction with the item $i$, otherwise $r_{ui}=0$.
Also, for a user $u$, $\mathcal{I}_{u}^{-} = \{ i \in \mathcal{I} | r_{ui} = 0\}$ denotes the set of unobserved items and $\mathcal{I}_{u}^{+} = \{ i \in \mathcal{I} | r_{ui} = 1\}$ denotes the set of interacted items.
A top-$K$ recommender system aims to find a recommendation list of unobserved items for each user.
To make the recommendation list, the system predicts the score $\hat{r}_{ui} = P(r_{ui}=1|u,i)$ for each item $i$ in $\mathcal{I}_{u}^{-}$ for each user $u$, then ranks the unobserved items according to their scores.

%% file: 4model.tex
\begin{figure*}[t]
\includegraphics[width=0.8\linewidth]{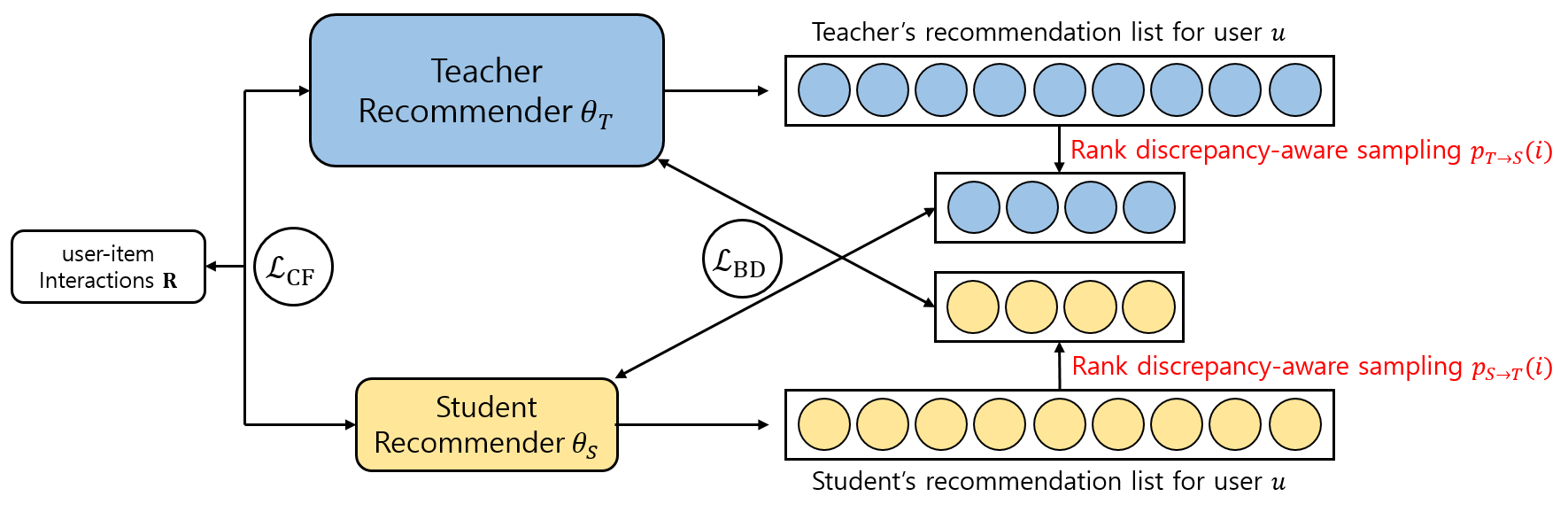}
\caption{Illustration of Bidirectional Distillation (BD) for top-$K$ recommender systems.}
\end{figure*}

We propose a novel \textbf{\underline{B}}idirectional \textbf{\underline{D}}istillation (BD) framework for top-$K$ recommender systems.
Within our proposed framework, both the teacher and the student transfer their knowledge to each other.
We first provide an overview of BD (Section 4.1).
Then, we formalize the distillation loss to transfer the knowledge between the two recommenders (Section 4.2).
We also propose \textit{rank discrepancy-aware sampling} scheme differently tailored for the teacher and the student to fully enhance each other (Section 4.3).
Lastly, we provide the details of the end-to-end training process within the proposed framework (Section 4.4).

\subsection{Overview}
Figure 2 shows an overview of BD.
Unlike the existing KD methods for RS, both the teacher and the student are trained simultaneously by using each other's knowledge (i.e., recommendation list) along with the binary training set (i.e., user-item interactions).
First, the teacher and the student produce the recommendation list for each user.
Second, BD decides what knowledge to be transferred for each distillation direction based on the rank discrepancy-aware sampling scheme.
Lastly, the teacher and the student are trained with the distillation loss along with the original collaborative filtering loss.
To summarize, each of them is  trained as follows:
\begin{equation}
\begin{split}
& \mathcal{L}_T(\theta_T)  = \mathcal{L}_{CF}(\theta_T) + \lambda_{S \rightarrow T} \cdot \mathcal{L}_{BD}(\theta_T; \theta_S), \\
& \mathcal{L}_S(\theta_S)  = \mathcal{L}_{CF}(\theta_S) + \lambda_{T \rightarrow S} \cdot \mathcal{L}_{BD}(\theta_S; \theta_T),
\end{split}
\end{equation}
where $T$ and $S$ denote the teacher and the student respectively, $\theta_*$ is the model parameters.
$\mathcal{L}_{CF}$ is the collaborative filtering loss depending on the base model which can be any existing recommender, and $\mathcal{L}_{BD}$ is the bidirectional distillation loss.
Lastly, $\lambda_{S \rightarrow T}$ and $\lambda_{T \rightarrow S}$ are the hyperparameters that control the effects of the distillation loss in each direction.

Within BD, the teacher and the student are collaboratively improved with each other based on their complementarity.
Also, during the training, the knowledge distilled between the teacher and the student gets gradually evolved along with the recommenders;
the improvement of the teacher leads to the acceleration of the student's learning, and the accelerated student again improves the teacher's learning.
Trained in the bidirectional way, both the teacher and the student are significantly improved compared to when being trained separately.
As a result, the student trained with BD outperforms the student model trained with the conventional distillation that relies on a pre-trained and fixed teacher.

\subsection{Distillation Loss}
We formalize the distillation loss that transfers the knowledge between the recommenders.
By following the original distillation loss that matches the class distributions of two classifiers for a given image \cite{kd15},
we design the distillation loss for the user $u$ as follows: 
\begin{equation}
\begin{aligned}
\mathcal{L}_{BD}(\theta_T; \theta_S) &=  \sum_{j \in \mathcal{RDS}_{S \rightarrow T}(\mathcal{I}_{u}^{-})} \mathcal{L}_{BCE}(\hat{r}^T_{uj}, \hat{r}^S_{uj})\\
\mathcal{L}_{BD}(\theta_S; \theta_T) &=  \sum_{j \in \mathcal{RDS}_{T \rightarrow S}(\mathcal{I}_{u}^{-})} \mathcal{L}_{BCE}(\hat{r}^S_{uj}, \hat{r}^T_{uj}),
\end{aligned}
\end{equation}
where $\mathcal{L}_{BCE}(p, q) = q \log p + (1-q) \log (1-p)$ is the binary cross-entropy loss,
$\hat{r}_{uj} = P(r_{uj}=1 | u,j)$ is the prediction of a recommender and $\mathcal{RDS}_{*}(\mathcal{I}_{u}^{-})$ is a set of the unobserved items sampled by the rank discrepancy-aware sampling.
$\hat{r}_{uj}$ is computed by $\sigma(z_{uj}/T)$
where $\sigma(\cdot)$ is the sigmoid function, $z_{uj}$ is the logit, and $T$ is the temperature that controls the smoothness.

Our distillation loss is a binary version of the original KD loss function.
Similar to the original KD loss transferring the knowledge of class probabilities, our loss transfers a user's potential positive and negative preferences on the unobserved items.
Specifically, in the binary training set, the unobserved interaction $r_{uj}$ is only labeled as “0”.
However, as mentioned earlier, it is ambiguous whether the user actually dislikes the item or potentially likes the item.
Through the distillation, each recommender can get the other's opinion of how likely (and unlikely) the user would be interested in the item, and such information helps the recommenders to better cope with the ambiguous nature of RS.

\subsection{Rank Discrepancy-aware Sampling}
We propose the rank discrepancy-aware sampling scheme that decides what knowledge to be transferred for each
distillation direction.
As we observed in Section 2, most of the items ranked highly by the teacher are already ranked highly by the student and vice versa.
Thus, the existing methods \cite{cd19,rd18} that simply choose the high-ranked items cannot give enough information to the other recommender.
Moreover, for effective bidirectional distillation, the performance gap between the teacher and the student should be carefully considered in deciding what knowledge to be transferred, as it is obvious that not all knowledge of the student is helpful to improve the teacher.

In this regard, we develop a sampling scheme based on the rank discrepancy of the teacher and the student, and tailor it differently for each distillation direction with consideration of their different capacities.
The underlying idea of the scheme is that each recommender can get informative knowledge by focusing on the items ranked highly by the other recommender, but ranked lowly by itself.
The sampling strategy for each distillation direction is defined as follows:

\vspace{3pt} 
\noindent
\textbf{Distillation from the teacher to the student.}
As the teacher has a much better overall performance than the student, the opinion of the teacher should be considered more reliable than that of the student in most cases.
Thus, for this direction of the distillation, we make the student follow the teacher's predictions on many rank-discrepant items.
Formally, for each user $u$, the probability of an item $i$ to be sampled is computed as follows:
\begin{equation}
p_{T \rightarrow S}(i) \propto tanh(\text{max}((rank_S(i) - rank_T(i)) \cdot \epsilon_t, 0)),
\end{equation}
where $rank_T(i)$ and $rank_S(i)$ denote the ranks assigned by the teacher and the student on the item $i$ for the user $u$ respectively, and $rank_{*}(i)=1$ is the highest ranking\footnote{we omit the superscript $u$ from $rank_{*}^{u}(i)$ for the simplicity.}.
We use a hyper-parameter $\epsilon_t$ $(> 0)$ to control the smoothness of the probability.
With this probability function, we sample the items ranked highly by the teacher but ranked lowly by the student.
Since $tanh(\cdot)$ is a saturated function, items with rank discrepancy above a particular threshold would be sampled almost uniformly.
As a result, the student learns the teacher's broad knowledge of most of the rank-discrepant items.

\vspace{3pt} \noindent \textbf{Distillation from the student to the teacher.}
As shown in Section 2, the teacher is not always superior to the student, especially for RS.
That is, the teacher can be also further improved by learning from the student.
However, at the same time, the large performance gap between the teacher and the student also needs to be considered for effective distillation.
For this direction of the distillation, we make the teacher follow the student's predictions on only a few selectively chosen rank-discrepant items.
The distinct probability function is defined as follows:
\begin{equation}
p_{S \rightarrow T}(i) \propto exp((rank_T(i) - rank_S(i)) \cdot \epsilon_e),
\end{equation}
where $\epsilon_e$ $(> 0)$ is a hyper-parameter to control the smoothness of the probability.
We use the exponential function to put particular emphasis on the items that have large rank discrepancies.
Therefore, this probability function enables the teacher to follow the student's predictions only on the rank-discrepant items that the student has very high confidence in.

\subsection{Model Training}

\begin{algorithm}[t]
\SetKwInOut{Input}{Input}
\SetKwInOut{Output}{Output}
\Input{Training data $\mathcal{D}$, the number of total epochs $e$, rank updating period $p$}
\Output{Teacher model $(\theta_T)$, Student model $(\theta_S)$}
Warm up $\theta_T$ and $\theta_S$ with only $\mathcal{L}_{CF}$\\
\For{$t=0,1,...,(e-1)$}{
\If{$t \text{ } \% \text{ } p == 0$}{
Teacher and Student update their recommendation lists
}
\For{$(u,i) \in \mathcal{D}$}{
\BlankLine
\tcc{Train Teacher}
Draw $\mathcal{RDS}_{S \rightarrow T}(\mathcal{I}_{u}^{-})$ with probability $p_{S \rightarrow T}(\cdot)$ \\
Compute $\mathcal{L}_{CF}(\theta_T)$ and $\mathcal{L}_{BD}(\theta_T; \theta_S)$ \\
Update $\theta_T$
\BlankLine
\tcc{Train Student}
Draw $\mathcal{RDS}_{T \rightarrow S}(\mathcal{I}_{u}^{-})$ with probability $p_{T \rightarrow S}(\cdot)$ \\
Compute $\mathcal{L}_{CF}(\theta_S)$ and $\mathcal{L}_{BD}(\theta_S; \theta_T)$ \\
Update $\theta_S$
}
}
\caption{Bidirectional Distillation Framework.}
\end{algorithm}

Algorithm 1 describes a pseudo code for the end-to-end training process within BD framework.
The training data $\mathcal{D}$ consists of observed interactions $(u,i)$.
First, the model parameters $\theta_T$ and $\theta_S$ are warmed up only with the collaborative filtering loss (line 1), as the predictions during the first few epochs are very unstable.
Second, we make the recommenders produce the recommendation lists for the subsequent sampling.
Since it is time-consuming to produce the recommendation lists every epoch, we conduct this step every $p$ epochs (line 3-4).
Next, we decide what knowledge to be transferred in each distillation direction via the rank discrepancy-aware sampling.
It is worth noting that the unobserved items sampled by the rank discrepancy-aware sampling can be used also for the collaborative filtering loss.
Finally, we compute the losses with the sampled items for the teacher and the student, respectively, and update the model parameters.

%% file: 5ex.tex
In this section, we validate our proposed framework on 9 experiment settings (3 real-world datasets $\times$ 3 base models).
We first introduce our experimental setup (Section 5.1).
Then, we provide a performance comparison supporting the superiority of the BD (Section 5.2).
We also provide two analyses for the in-depth understanding of BD (Section 5.3, 5.4).
Lastly, We provide an ablation study to verify the effectiveness of rank discrepancy-aware sampling (Section 5.5) and analyses for important hyperparameters of BD (Section 5.6).

\subsection{Experimental Setup}
\subsubsection{Datasets}
We use three real-world datasets: CiteULike\footnote{https://github.com/changun/CollMetric/tree/master/citeulike-t} \cite{CUL13}, Foursquare\footnote{https://sites.google.com/site/yangdingqi/home/foursquare-dataset} (Tokyo Check-in) \cite{FS14} and Yelp\footnote{https://github.com/hexiangnan/sigir16-eals/blob/master/data/yelp.rating} \cite{yelp16}.
We only keep users who have at least five ratings for CiteULike and Foursquare, ten ratings for Yelp as done in \cite{ncf17, bpr09}.
Data statistics after the preprocessing are presented in Table 1.
We also report the experimental results on ML100K and AMusic, which are used for CD \cite{cd19}, in Appendix for the direct comparison.
\begin{table}[t]
  \renewcommand{\arraystretch}{0.7}
  \caption{Data Statistics}
  \begin{tabular}{ccccc}
    \toprule
    Dataset & \#Users & \#Items & \#Ratings & Sparsity \\
    \midrule
    CiteULike & 5,219 & 25,187 & 130,788 & 99.90\% \\
    Foursquare & 2,293 & 61,858 & 537,167 & 99.62\% \\
    Yelp & 25,677 & 25,815 & 730,623 & 99.89\% \\
    \bottomrule
  \end{tabular}
\end{table}

\subsubsection{Evaluation Protocol and Metrics}
We adopt the widely used \textit{leave-one-out} evaluation protocol.
For each user, we hold out the last interacted item for testing and the second last interacted item for validation as done in \cite{cd19, ncf17}.
If there is no timestamp in the dataset, we randomly take two observed items for each user.
Then, we evaluate how well each method can rank the test item higher than all the unobserved items for each user (i.e., $\mathcal{I}_u^{-}$).
Note that instead of randomly choosing a predefined number of candidates (e.g., 99), we adopt the full-ranking evaluation that uses all the unobserved items as candidates.
Although it is time-consuming, it enables a more thorough evaluation compared to using random candidates \cite{fullrank20, cd19}.

As we focus on top-$K$ recommendation for implicit feedback, we employ two widely used metrics for evaluating the ranking performance of recommenders: Hit Ratio (H@$K$) \cite{hr16} and Normalized Discounted Cumulative Gain (N@$K$) \cite{NDCG02}.
H@$K$ measures whether the test item is present in the top-$K$ list and N@$K$ assigns a higher score
to the hits at higher rankings in the top-$K$ list.
We compute those two metrics for each user, then compute the average score.
Lastly, we report the average value of five independent runs for all methods.

\subsubsection{Base Models}
BD is a model-agnostic framework applicable for any top-$K$ RS.
We validate BD with three base models that have different model architectures and learning strategies.
Specifically, we choose two widely used deep learning models and one latent factor model as follows:
\begin{itemize}
    \item \textbf{NeuMF \cite{ncf17}}: A deep recommender that adopts Matrix Factorization (MF) and Multi-Layer Perceptron (MLP) to capture complex and non-linear user-item relationships. 
    NeuMF uses the point-wise loss function for the optimization.
    \item \textbf{CDAE \cite{cdae16}}: 
    A deep recommender that adopts Denoising Autoencoders (DAE) \cite{dae08} for the collaborative filtering. 
    CDAE uses the point-wise loss function for the optimization.
    \item \textbf{BPR \cite{bpr09}}: A learing-to-rank recommender that adopts MF \cite{mf09} to model the user-item interaction.
    BPR uses the pair-wise loss function for the optimization under the assumption that observed items are more preferred than unobserved items.
\end{itemize}

\input{5exzmaintable}

\subsubsection{Methods Compared}
The proposed framework is compared with the state-of-the-art KD methods for top-$K$ RS.
\begin{itemize}
    \item \textbf{Ranking Distillation (RD) \cite{rd18}}: A pioneering KD method for top-$K$ RS.
    RD makes the student give high scores on top-ranked items by the teacher.
    \item \textbf{Collaborative Distillation (CD) \cite{cd19}}: A state-of-the-art KD method for top-$K$ RS. 
    CD makes the student imitate the teacher's scores on the items ranked highly by the teacher.
\end{itemize}

\subsubsection{Implementation Details}
For all the base models and baselines, we use PyTorch \cite{pytorch19} for the implementation.
For each dataset, hyperparameters are tuned by using grid searches on the validation set. 
We use Adam optimizer \cite{adam14} with L2 regularization and the learning rate is chosen from $\{$0.00001, 0.0001, 0.001, 0.002$\}$, and we set the batch size as 128.
For NeuMF, we use 2-layer MLP for the network.
For CDAE, we use 2-layer MLP for the encoder and the decoder, and the dropout ratio is set to 0.5.
The number of negative samples is set to 1 for NeuMF and BPR, $5*|\mathcal{I}_{u}^{+}|$ for CDAE as suggested in the original paper \cite{cdae16}.

For the distillation, we adopt as many learning parameters as possible for the teacher model until the ranking performance is no longer increased on each dataset.
Then, we build the student model by employing only one-tenth of the learning parameters used by the teacher.
The number of model parameters of each base model is reported in Table 3.
For KD competitors (i.e., RD, CD), $\lambda_{KD}$ is chosen from $\{$0.01, 0.1, 0.5, 1$\}$, the number of items sampled for the distillation is chosen from $\{$10, 15, 20, 30, 40, 50$\}$, and the temperature $T$ for logits is chosen from $\{$1, 1.5, 2$\}$.
For other hyperparameters, we use the values recommended from the public implementation and the original papers \cite{rd18, cd19}.
For BD, $\lambda_{T \rightarrow S}$ and $\lambda_{S \rightarrow T}$ are set to 0.5, the number of items sampled by rank discrepancy-aware sampling is chosen from $\{$1, 5, 10$\}$, $\epsilon_t$ is chosen from $\{10^{-2}, 10^{-3}, 10^{-4}, 10^{-5}\}$, $\epsilon_e$ is chosen from $\{10^{-3}, 10^{-4}, 10^{-5}\}$, the temperature $T$ for logits is chosen from $\{$2, 5, 10$\}$ and the rank updating period $p$ (in Algorithm 1) is set to 10.

\subsection{Performance Comparison}
Table 2 shows the recommendation performance of each KD method on three real-world datasets and three different base models.
In Table 2, "Teacher" and "Student" indicate the base models trained separately without any distillation technique, "BD-Teacher" and "BD-Student" are the teacher and the student trained simultaneously with BD.
"CD" and "RD" denote the student trained with CD and RD, respectively.
We analyze the experimental results from various perspectives.

\input{5exzmodelparams}

\begin{figure*}[t]
\begin{subfigure}[t]{0.495\linewidth}
    \includegraphics[width=0.5\linewidth]{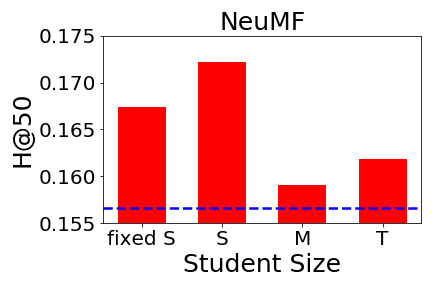}
    \includegraphics[width=0.5\linewidth]{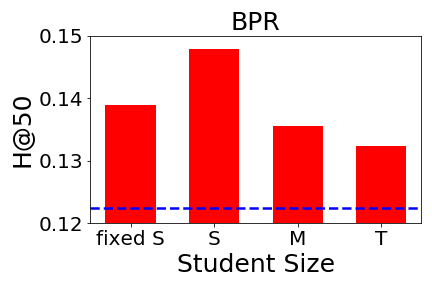}
    \caption{Performance of teacher with student of varying size}
\end{subfigure} 
\begin{subfigure}[t]{0.495\linewidth}
    \includegraphics[width=0.5\linewidth]{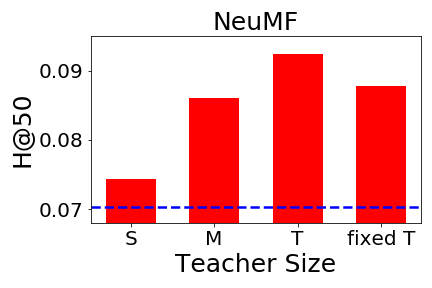}
    \includegraphics[width=0.5\linewidth]{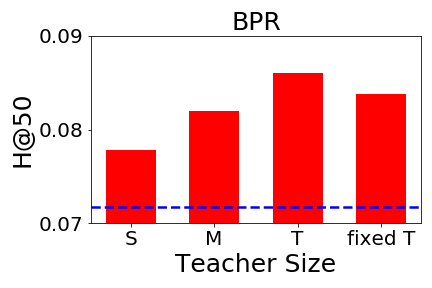}
    \caption{Performance of student with teacher of varying size}
\end{subfigure}
\caption{Effect of the model size on BD. The blue dashed line indicates the performance of the model trained separately.}
\end{figure*}

We first observe that the teacher recommender is consistently improved by the knowledge transferred from the student with BD by up to 20.79\% (for H@50 on CiteULike).
This result verifies our claim that the knowledge of the student also could be useful for improving the teacher.
Also, this result strongly indicates that the distillation process of BD effectively resolves the challenge of a large performance gap, and successfully transfer the informative knowledge from the student.
In specific, the ranking discrepancy-aware sampling enables the teacher to focus on the items that the student has very high confidence in.
This strategy can minimize the adverse effects from the performance gap, making the teacher be further improved based on the complementary knowledge from the student.

Second, the student recommender is significantly improved by the knowledge transferred from the teacher with BD by up to 39.88\% (for N@50 on CiteULike).
Especially, the student trained with BD considerably outperforms the student trained with the existing KD methods (i.e., RD, CD) by up to 13.94\% (for H@100 on Yelp).
The superiority of BD comes from the two contributions;
BD makes the student follow the teacher’s predictions on the rank-discrepant items which are more informative than the merely high-ranked items.
Also, within BD, the improvement of the teacher leads to the acceleration of the student’s learning, and the accelerated student again improves the teacher’s learning.
With better guidance from the improved teacher, the student with BD can achieve superior performance than the student with the existing KD methods.
To verify the effectiveness of each contribution, we conduct in-depth analyses in the next sections.

Lastly, Table 3 shows the model size and online inference efficiency for each base model.
For making the inferences, we use PyTorch with CUDA on GTX Titan X GPU and Intel i7-7820X CPU.
As the student has only one-tenth of the learning parameters used by the teacher, the student requires less computational costs, thus achieves lower inference latency.
Moreover, the student trained with BD shows comparable or even better recommendation performance to the teacher (e.g., NeuMF and CDAE on Foursquare).
On CiteULike, we observe that the student achieves comparable performance by employing 20$\sim$30\% of learning parameters used by the teacher.
These results show that BD can be effectively adopted to train a small but powerful recommender.
Moreover, as mentioned earlier, BD is also applicable in setting where there is no constraint on the model size or inference time.
Specifically, BD can be adopted to maximize the performance of a large recommender with numerous learning parameters (i.e., the teacher).

\subsection{Model Size Analysis}
We control the size of the teacher and the student to analyze the effects of the capacity gap between two recommenders on BD.
We report the results of a deep model and a latent factor model (i.e., NeuMF and BPR) on CiteULike.
Figure 3a shows the performance of the teacher trained with the student of varying sizes and Figure 3b shows the performance of the student trained with the teacher of varying sizes.
“S” indicates the size of the student (10\% of the teacher), “M” indicates the medium size (50\% of the teacher), and “T” indicates the size of the teacher.
Also, “fixed S” refers to the pre-trained student recommender with the size of "S", “fixed T” refers to the pre-trained teacher recommender with the size of "T".
Note that "fixed S/T" are not updated during the training (i.e., unidirectional knowledge distillation).

First, we observe that both the teacher and the student achieves the greatest performance gain when the capacity gap between two recommenders is largest;
the teacher shows the best performance with the smallest student (i.e., S in Fig. 3a) and the student performs best with the largest teacher (i.e., T in Fig. 3b).
As mentioned in Section 2, the teacher and the student have complementarity as some user-item relationships can be better captured without expensive computations.
In this regard, such complementarity can be maximized when the capacity gap between the two recommenders is large enough.
This result supports our claim that the performance gain comes from the different but reciprocal knowledge of two recommenders.
Also, it is worth noting that there are still performance improvements when two recommenders have identical sizes (i.e., T in Fig. 3a, S in Fig. 3b).
This can be understood as a kind of self-distillation effect \cite{born18, self19} when the teacher has the same size as the student.
Although they have very similar kinds of knowledge, they can still regularize each other, preventing its counterpart from being overfitted to a few observed interactions.

Second, the teacher is more improved when it is trained along with the learning student than when it is trained with the fixed student (i.e., S vs. fixed S in Fig. 3a).
Similarly, the student is more improved when it is trained together with the learning teacher than when it is trained with the fixed teacher (i.e., T vs. fixed T in Fig. 3b).
In the unidirectional distillation, the pre-trained recommender (i.e., fixed S/T) is no longer improved, thus always conveys the same knowledge during the training.
On the contrary, within BD, both recommenders are trained together, thus the knowledge distilled between the recommenders gets gradually evolved as they are improved during the training.
As a result, each recommender can be further improved based on the evolved knowledge of its counterparts.
This result again shows the superiority of our bidirectional distillation over the unidirectional distillation of the existing methods.

\subsection{Synchronization Analysis}
\begin{table}[t]
  \renewcommand{\arraystretch}{0.8}
  \caption{Average Rank Difference before and after the training with BD.}
  \begin{tabular}{ccccc}
    \toprule
    \multicolumn{2}{c}{Base Model} & CiteULike & Foursquare & Yelp\\
    \midrule
    \multirow{2}{*}{NeuMF} & before BD & 0.1944 & 0.1020 & 0.0918\\
    & after BD & 0.1323 & 0.0957 & 0.0824\\
    \midrule
    \multirow{2}{*}{CDAE} & before BD & 0.1190 & 0.0963 & 0.0544\\
    & after BD & 0.0871 & 0.0883 & 0.0472\\
    \midrule
    \multirow{2}{*}{BPR} & before BD & 0.1380 & 0.1180 & 0.0560\\
    & after BD & 0.1064 & 0.1082 & 0.0511\\
    \bottomrule
  \end{tabular}
\end{table}

\begin{figure}[t]
\begin{subfigure}[t]{\linewidth}
    \includegraphics[width=0.5\linewidth]{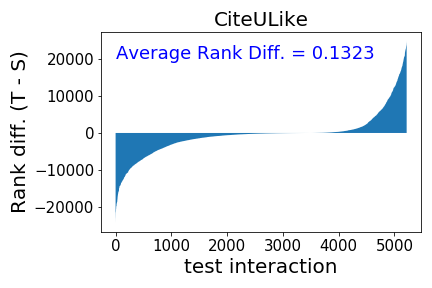}
    \includegraphics[width=0.5\linewidth]{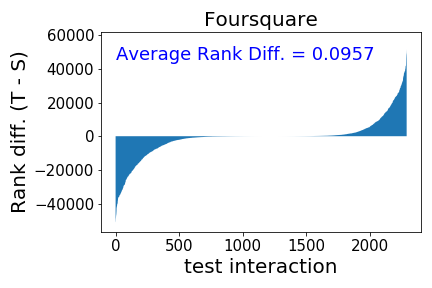}
\end{subfigure} 
\caption{Rank difference after the training with BD.}
\end{figure}

We perform in-depth analysis to investigate how well the teacher and the student learn each other's complementary knowledge within the proposed framework.
Specifically, we evaluate how much synchronized the two recommenders are within BD, which shows that they are improved based on each other's complementarity.
To quantify the degree of the synchronization, we define (normalized) Average Rank Difference as follows:
\begin{equation}
\text{Average Rank Diff.} = \frac{1}{n \cdot m} \sum_{\mathcal{D}_{test}} \abs{\text{Rank Diff.}^{u}(i)},
\end{equation}
where $n$ and $m$ are the numbers of users and items, respectively, and $\mathcal{D}_{test}$ is the test set that contains the held-out observed interaction for each user.
The rank difference ($\text{Rank Diff.}^{u}(i)$) is defined in Equation 1.

Table 4 shows the change in the average rank difference after the training with BD.
We observe that the average rank difference gets consistently decreased on all the datasets.
This result indicates that the teacher and the student get synchronized by transferring their knowledge to each other.
Figure 4 shows the rank difference between the teacher and the student after the training with BD.
We adopt NeuMF as the base model as done in Section 2.
Note that the (normalized) average rank difference is proportional to the extent of the blue area.
We observe that the blue area in Figure 4 shrinks compared to that in Figure 1 after the training with BD.
Interestingly, we observe that the teacher is significantly improved on CiteULike dataset (by up to 20.79\%) which has the largest average rank difference change before and after the training with BD.
The large change indicates that the two recommenders get synchronized well during the training, which leads to significant improvements based on each other's knowledge.

\subsection{Sampling Scheme Analysis}
We examine the effects of diverse sampling schemes on the performance of BD to verify the superiority of the proposed sampling scheme.
Note that the schemes decide what knowledge to be distilled within BD.
We compare five different sampling schemes as follows: 
1) Rank discrepancy-aware sampling, 
2) Rank-aware sampling \cite{cd19},
3) Top-$N$ selection \cite{rd18},
4) Uniform sampling, 
5) \textit{Swapped} rank discrepancy-aware sampling.

The rank discrepancy-aware sampling, which is our proposed scheme, focuses on the rank-discrepant items between the teacher and the student.
On the other hand, the rank-aware sampling (adopted in CD) and top-$N$ selection (adopted in RD) focus on the items ranked highly by one recommender.
The uniform sampling randomly chooses items from the entire recommendation list.
Finally, the swapped rank discrepancy-aware sampling, which is the ablation of the proposed scheme, swaps the sampling probability function of each distillation direction;
we make the teacher follow the student on most of the rank-discrepant items (with \textit{tanh}), and make the student imitate only a few predictions of the teacher (with \textit{exp}).

\begin{table}[t]
\small
  \renewcommand{\arraystretch}{1}
  \caption{Recommendation performance (H@50) of BD with different sampling schemes on CiteULike. Numbers in bold face are the best results.}
  \begin{tabular}{cc cc}
    \toprule
    Base Model & Sampling Scheme & Teacher & Student\\
    \midrule
    \multirow{5}{*}{NeuMF} & Rank discrepancy-aware & \textbf{0.1722} & \textbf{0.0924} \\
    & Rank-aware \cite{cd19} & 0.1617 & 0.0812 \\
    & Top-$N$ selection \cite{rd18} & 0.1480 & 0.0766 \\
    & Uniform & 0.1590 & 0.0726 \\
    & Swapped rank discrepancy-aware & 0.1512 & 0.0747 \\
    \midrule
    \multirow{5}{*}{CDAE} & Rank discrepancy-aware & \textbf{0.1983} & \textbf{0.0943} \\
    & Rank-aware \cite{cd19} & 0.1818 & 0.0891 \\
    & Top-$N$ selection \cite{rd18} & 0.1757 & 0.0870 \\
    & Uniform & 0.1788 & 0.0819 \\
    & Swapped rank discrepancy-aware & 0.1733 & 0.0851 \\
    \midrule
    \multirow{5}{*}{BPR} & Rank discrepancy-aware & \textbf{0.1479} & \textbf{0.0853} \\
    & Rank-aware \cite{cd19} & 0.1367 & 0.0805 \\
    & Top-$N$ selection \cite{rd18} & 0.1264 & 0.0772 \\
    & Uniform & 0.1306 & 0.0749 \\
    & Swapped rank discrepancy-aware & 0.1281 & 0.0803 \\
    \bottomrule
  \end{tabular}
\end{table}

Table 5 shows the performance of BD with different sampling schemes.
We first observe that the proposed scheme achieves the best result both for the teacher and the student.
This result verifies the effectiveness of two components of the proposed scheme: 1) sampling based on the rank discrepancy, 2) probability functions differently designed for each distillation direction.
Specifically, we can see the effectiveness of rank discrepant-based sampling by comparing our sampling scheme with rank-aware sampling.
As observed in Section 2, the items merely ranked highly by the teacher may be not informative enough to fully enhance the student.
With the proposed scheme, BD focuses on the rank-discrepant items, thus can further improve the recommenders.
Also, we observe that swapping the probability function of each distillation direction (i.e., the swapped rank discrepancy-aware) significantly degrades the performance.
This result strongly indicates that each probability function is well designed to effectively cope with the large performance gap of two recommenders.

\subsection{Hyperparameter Analysis}
In this section, we provide thorough analyses that examine the effects of important hyperparameters on BD.
For the sake of the space, we report the results of NeuMF on CiteULike dataset.
We observe similar tendencies with other base models and datasets.

\begin{figure}[t]
\includegraphics[width=\linewidth, scale=1.7]{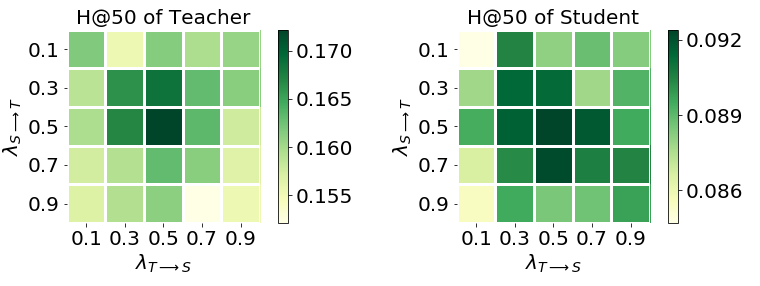}
\caption{Recommendation performance (H@50) of teacher and student with varying lambda.}
\end{figure}

\begin{figure}[t]
\begin{subfigure}[t]{\linewidth}
    \includegraphics[width=0.5\linewidth]{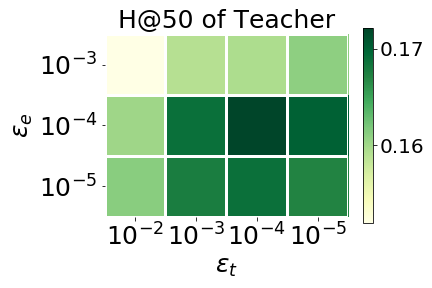}
    \includegraphics[width=0.5\linewidth]{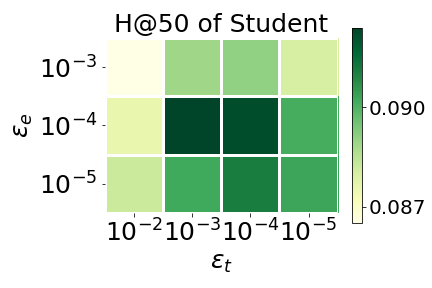}
\end{subfigure} 
\caption{Recommendation performance (H@50) of teacher and student with varying smoothing factor.}
\end{figure}

First, Figure 5 shows the performance of the teacher and the student with varying $\lambda_{T \rightarrow S}$ and $\lambda_{S \rightarrow T}$ which control the effects of the distillation losses.
The best performance is achieved when both $\lambda_{T \rightarrow S}$ and $\lambda_{S \rightarrow T}$ are around 0.5.
Also, we observe that both the recommenders are considerably improved when $\lambda_{T \rightarrow S}$ and $\lambda_{S \rightarrow T}$ have similar values (i.e., diagonal entries).
Moreover, we observe that the performance of the student is more robust with respect to $\lambda$ than that of the teacher.
We believe that this is because the student has a lower overall performance than the teacher, thus can easily take advantage of the teacher in broad settings.

\begin{figure}[t]
\begin{subfigure}[t]{\linewidth}
    \includegraphics[width=0.5\linewidth]{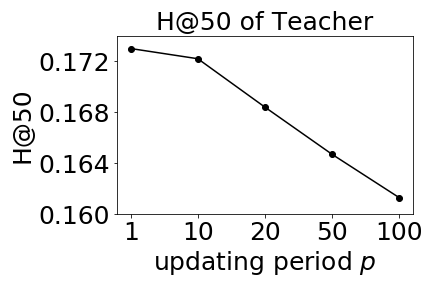}
    \includegraphics[width=0.5\linewidth]{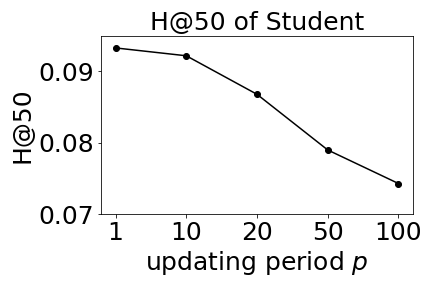}
\end{subfigure} 
\caption{Recommendation performance (H@50) of teacher and student with varying updating period.}
\end{figure}

Second, Figure 6 shows the performance of the teacher and the student with varying $\epsilon_t$ and $\epsilon_e$ which control the smoothness of the probability functions in rank discrepancy-aware sampling.
When $\epsilon_t$ and $\epsilon_e$ are small, the probability functions become smooth.
The best performance is achieved when both parameters are around $10^{-4}$.
When $\epsilon_e$ is bigger than $10^{-3}$, the probability $p_{S \rightarrow T}(\cdot)$ gets too sharp.
Thus, the teacher cannot fully learn the student's complementary knowledge, which leads to degraded performance.

Lastly, Figure 7 shows the performance of the teacher and the student with varying rank updating period $p$ (in Algorithm 1).
Since it is time-consuming to generate the recommendation list every epoch, we update the recommendation list of the two models every $p$ epoch.
In this paper, we use $p=10$ which shows the comparable performance to the upper-bound ($p=1$).

%% file: 5exzmaintable.tex
\begin{table*}[t]
 \small
 \renewcommand{\arraystretch}{0.7}
 \caption{Performance comparison. \textit{Improv.T} denotes the improvement of the teacher over the original teacher model ("Teacher" in this table), \textit{Improv.B} denotes the improvement of the student over the best competitive method and \textit{Improv.S} denotes the improvement of the student over the original student model ("Student" in this table). $*$ and $**$ indicate $p \leq $ 0.05 and $p \leq $ 0.01 for the paired t-test of BD vs. the best competitor.}
  \begin{tabular}{cl cccc | cccc | cccc}
    \toprule 
    \multirow{2}{*}{Base Model} & \multirow{2}{*}{Method} & \multicolumn{4}{c}{CiteULike} & \multicolumn{4}{c}{Foursquare} & \multicolumn{4}{c}{Yelp}\\
     & & H@50 & H@100 & N@50 & N@100 & H@50 & H@100 & N@50 & N@100 & H@50 & H@100 & N@50 & N@100 \\
    \bottomrule
     & Teacher & 0.1566 & 0.2272 & 0.0385 & 0.0487 & 0.1612 & 0.2066 & 0.0615 & 0.0685 & 0.0782 & 0.1341 & 0.0200 & 0.0290 \\
     & BD-Teacher & 0.1722 & 0.2443 & 0.0431 & 0.0548 & 0.1704 & 0.2232 & 0.0653 & 0.0737 & 0.0836 & 0.1473 & 0.0217 & 0.0319 \\
     & BD-Student & 0.0924 & 0.1425 & 0.0242 & 0.0325 & 0.1657 & 0.2181 & 0.0637 & 0.0721 & 0.0779 & 0.1315 & 0.0203 & 0.0289  \\
     & CD & 0.0820 & 0.1318 & 0.0227 & 0.0307 & 0.1548 & 0.2063 & 0.0592 & 0.0677 & 0.0712 & 0.1225 & 0.0189 & 0.0263 \\
    NeuMF & RD & 0.0755 & 0.1242 & 0.0197 & 0.0268 & 0.1442 & 0.1819 & 0.0547 & 0.0611 & 0.0662 & 0.1134 & 0.0161 & 0.0237 \\
     & Student & 0.0703 & 0.1167 & 0.0173 & 0.0248 & 0.1264 & 0.1674 & 0.0512 & 0.0578 & 0.0615 & 0.1018 & 0.0153 & 0.0221 \\
    \cmidrule{2-14}
     & \textit{Improv.T} & 9.96\% & 7.53\% & 11.95\% & 12.53\% & 5.73\% & 8.03\% & 6.21\% & 7.59\% & 6.91\% & 9.84\% & 8.50\% & 10.00\% \\
     & \textit{Improv.B} & 12.68\%{**} & 8.16\%{**} & 6.84\%{**} & 5.87\%{**} & 7.05\%{**} & 5.74\%{**} & 7.53\%{**} & 6.50\%{**} & 9.41\%{**} & 7.35\%{**} & 7.41\%{**} & 9.89\%{**} \\
     & \textit{Improv.S} & 31.44\% & 22.11\% & 39.88\% & 30.85\% & 31.11\% & 30.31\% & 24.34\% & 24.74\% & 26.67\% & 29.17\% & 32.68\% & 30.77\% \\
    \midrule
     & Teacher & 0.1710 & 0.2445 & 0.0492 & 0.0611 & 0.1653 & 0.2281 & 0.0650 & 0.0743 & 0.0894 & 0.1523 & 0.0229 & 0.0331 \\
     & BD-Teacher & 0.1983 & 0.2702 & 0.0588 & 0.0674 & 0.1740 & 0.2368 & 0.0680 & 0.0766 & 0.0993 & 0.1681 & 0.0255 & 0.0366 \\
     & BD-Student & 0.0943 & 0.1470 & 0.0269 & 0.0355 & 0.1721 & 0.2242 & 0.0614 & 0.0709 & 0.0745 & 0.1323 & 0.0191 & 0.0268 \\
     & CD & 0.0861 & 0.1332 & 0.0250 & 0.0324 & 0.1629 & 0.2104 & 0.0553 & 0.0651 & 0.0667 & 0.1183 & 0.0176 & 0.0236 \\
    CDAE & RD & 0.0848 & 0.1266 & 0.0241 & 0.0318 & 0.1670 & 0.2146 & 0.0572 & 0.0675 & 0.0657 & 0.1118 & 0.0157 & 0.0231 \\
     & Student & 0.0724 & 0.1090 & 0.0198 & 0.0257 & 0.1491 & 0.2041 & 0.0552 & 0.0641 & 0.0613 & 0.1061 & 0.0151 & 0.0218 \\
    \cmidrule{2-14}
     & \textit{Improv.T} & 15.96\% & 10.51\% & 13.41\% & 10.31\% & 5.26\% & 3.81\% & 4.62\% & 3.10\% & 11.07\% & 10.37\% & 11.35\% & 10.57\% \\
     & \textit{Improv.B} & 9.52\%{**} & 10.36\%{**} & 7.60\%{**} & 9.57\%{**} & 3.05\%{*} & 4.47\%{*} & 7.34\%{**} & 5.04\%{**} & 11.86\%{**} & 11.83\%{**} & 8.52\%{**} & 13.56\%{**} \\
     & \textit{Improv.S} & 30.25\% & 34.86\% & 35.86\% & 38.13\% & 15.43\% & 9.85\% & 11.23\% & 10.61\% & 21.53\% & 24.69\% & 26.49\% & 22.94\% \\
    \midrule
     & Teacher & 0.1224 & 0.1892 & 0.0328 & 0.0415 & 0.1746 & 0.2307 & 0.0662 & 0.0764 & 0.0896 & 0.1383 & 0.0231 & 0.0291 \\
     & BD-Teacher & 0.1479 & 0.2144 & 0.0390 & 0.0498 & 0.1834 & 0.2449 & 0.0713 & 0.0812 & 0.0946 & 0.1488 & 0.0245 & 0.0302 \\
     & BD-Student & 0.0853 & 0.1378 & 0.0213 & 0.0298 & 0.1611 & 0.1991 & 0.0610 & 0.0687 & 0.0733 & 0.1283 & 0.0171 & 0.0243 \\
     & CD & 0.0803 & 0.1271 & 0.0201 & 0.0277 & 0.1515 & 0.1851 & 0.0567 & 0.0622 & 0.0660 & 0.1137 & 0.0144 & 0.0221 \\
    BPR & RD & 0.0782 & 0.1234 & 0.0185 & 0.0259 & 0.1404 & 0.1696 & 0.0519 & 0.0574 & 0.0661 & 0.1126 & 0.0155 & 0.0215 \\
     & Student & 0.0717 & 0.1217 & 0.0171 & 0.0251 & 0.1252 & 0.1618 & 0.0469 & 0.0528 & 0.0559 & 0.1039 & 0.0127 & 0.0199 \\
    \cmidrule{2-14}
     & \textit{Improv.T} & 20.79\% & 13.32\% & 18.90\% & 19.88\% & 5.04\% & 6.13\% & 7.63\% & 6.22\% & 5.58\% & 7.59\% & 6.06\% & 3.78\% \\
     & \textit{Improv.B} & 6.16\%{**} & 8.38\%{**} & 5.97\%{**} & 7.40\%{**} & 6.34\%{**} & 7.56\%{**} & 7.58\%{**} & 10.45\%{**} & 10.89\%{**} & 13.94\%{**} & 10.32\%{**} & 13.02\%{**} \\
     & \textit{Improv.S} & 18.90\% & 13.19\% & 24.56\% & 18.53\% & 28.67\% & 23.05\% & 30.06\% & 30.11\% & 31.13\% & 23.48\% & 34.65\% & 22.11\% \\
    \bottomrule
  \end{tabular}
\end{table*}

%% file: 5exzmodelparams.tex
\begin{table}[t]
\small
\setlength{\tabcolsep}{4pt}
  \renewcommand{\arraystretch}{0.8}
  \caption{Model size and online inference efficiency. Time denotes the wall time used for generating recommendation list for every user.}
  \begin{tabular}{c cc cc cc}
    \toprule
    \multirow{2}{*}{Base Model} & \multicolumn{2}{c}{CiteULike} & \multicolumn{2}{c}{Foursquare} & \multicolumn{2}{c}{Yelp}\\
     & \#Params. & Time & \#Params. & Time & \#Params. & Time \\
    \midrule
    NeuMF-T & 3.05M & 17.39s & 6.42M & 14.77s & 5.15M & 85.12s \\
    NeuMF-S & 0.30M & 14.12s & 0.64M & 7.99s & 0.51M & 69.53s \\
    \midrule
    CDAE-T & 2.80M & 14.73s & 6.36M & 7.95s & 3.89M & 64.45s \\
    CDAE-S & 0.30M & 10.35s & 0.69M & 5.78s & 0.41M & 53.39s \\
    \midrule
    BPR-T & 1.52M & 9.11s & 3.21M & 7.71s & 2.57M & 46.83s \\
    BPR-S & 0.15M & 7.94s & 0.32M & 5.39s & 0.26M & 40.49s \\
    \bottomrule
  \end{tabular}
\end{table}

%% file: 6rw.tex
\noindent \textbf{Reducing inference latency of RS.}
Several methods have been proposed for reducing the model size and inference time of recommender systems.
First, a few work adopt discretization techniques to reduce the size of recommenders \cite{discreterec1, discreterec2, discreteAAAI, binary12, discreterec3}.
They learn discrete representations of users and items to make portable recommenders and successfully reduce the model size.
However, their recommendation performance is highly limited due to the restricted capability and thus the loss of recommendation performance is unavoidable \cite{cd19}.
Second, several methods try to accelerate the inference phase by adopting model-dependent techniques \cite{treeRS, inference15, inference17}.
For example, the order-preserving transformations \cite{treeRS} and the pruning techniques \cite{inference15, inference17} have been adopted to reduce the computational costs of the inner product.
Although they can reduce the inference latency, they are applicable only to specific models (e.g., inner product-based models).

To tackle this challenge, a few recent methods \cite{rd18, cd19, DERRD} have adopted knowledge distillation to RS.
KD is a model-agnostic strategy and we can employ any recommender system as the base model. 
RD \cite{rd18} firstly proposes a KD method that makes the student give high scores on the top-ranked items of the teacher’s recommendation list.
Similarly, CD \cite{cd19} makes the student imitate the teacher’s prediction scores with particular emphasis on the items ranked highly by the teacher.
The most recent work, RRD \cite{DERRD}, formulates the distillation process as a relaxed ranking matching problem between the ranking list of the teacher and that of the student.
Since it is daunting for the small student to learn all the prediction results from the large teacher, they focus on the high-ranked items, which can affect the top-$K$ recommendation performance \cite{DERRD}. 
By using such supplementary supervisions from the teacher, they have successfully improved the performance of the student. 
However, they have some critical limitations in that 
1) they rely on the unidirectional distillation, 
2) the high-ranked items are not informative enough.
These limitations are thoroughly analyzed and resolved in this work. 

\vspace{3pt} \noindent \textbf{Training multiple models together.}
There have been successful attempts to train multiple models simultaneously for better generalization performance in computer vision and natural language processing \cite{dml18, collaborativelearning18, dual16}.
In computer vision, Deep Mutual Learning (DML) \cite{dml18} trains a cohort of multiple classifiers simultaneously in a peer-teaching scenario where each classifier is trained to follow the predictions of the other classifiers along with its original loss.
Since each classifier starts from a different initial condition, each classifier's predicted probabilities of the next most likely class vary.
DML claims that those secondary quantities provide extra information to the other classifiers that can help them to converge to a more robust minima.
Also, Collaborative Learning \cite{collaborativelearning18} trains several classifier heads of the same network simultaneously by sharing intermediate-level representations.
Collaborative Learning argues that the consensus of multiple views from different heads on the same data provides both supplementary information and regularization to each classifier head.
In natural language processing, Dual Learning \cite{dual16} proposes a learning mechanism that two machine translators teach each other to reduce the costs of human labeling.
Pointing out that the machine translation can be considered as a dual-task forming a closed loop (e.g., English-to-French/ French-to-English), 
Dual Learning generates informative feedback by transferring their outputs to each other and achieves considerable performance improvements without the involvement of a human labeler.
Although the aforementioned methods have effectively increased performance by training multiple models simultaneously, they mostly employ models with the same sizes focusing only on achieving better generalization based on the consensus of various views.

In this paper, we propose a novel Bidirectional Distillation framework whereby two recommenders of different sizes are trained together by transferring their knowledge to each other.
Pointing out that the large recommender and the small recommender have different but complementary knowledge, we design an effective distillation mechanism that considers their huge capacity gap.

%% file: 7con.tex
We propose a novel Bidirectional Distillation framework for top-$K$ recommender systems whereby the teacher and the student are collaboratively improved with each other during the training.
Within our proposed framework, both the teacher and the student transfer their knowledge to each other with the distillation loss.
Also, our framework considers the capacity gap between the teacher and the student with differently tailored \textit{rank discrepancy-aware sampling}.
Our extensive experiments on real-world datasets show that BD significantly outperforms the state-of-the-art competitors in terms of the student recommender.
Furthermore, the teacher model also gets the benefit of the student and performs better than when being trained separately.
We also provide analyses for the in-depth understanding of BD and verifying the effectiveness of each proposed component.

%% file: 8app.tex
\section{Appendix}
In this section, we report the experimental results on ML100K and AMusic, which are used for CD \cite{cd19}, for the direct comparison with CD.
We do not include this result in our main table, because we consider those datasets are relatively small to simulate the real-world evaluation (ML100K has only 943 users and 1682 items).
We adopt CDAE as the base model and we employ 2-layer MLP for the encoder and the decoder of CDAE.
Experimental results on their experimental settings are as follows:
\begin{table}[h]
\small
  \caption{Performance comparison with CD.}
  \begin{tabular}{c cc cc}
    \toprule
    \multirow{2}{*}{Model} & \multicolumn{2}{c}{ML100K} & \multicolumn{2}{c}{AMusic} \\
     & H@50 & N@50 & H@50 & N@50 \\
    \midrule
    Teacher & 0.3966 & 0.1259 & 0.1748 & 0.0533 \\
    BD-Teacher & 0.4317 & 0.1442 & 0.1936 & 0.0607 \\
    BD-Student & 0.4111 & 0.1321 & 0.1723 & 0.0545 \\
    CD & 0.3786 & 0.1232 & 0.1650 & 0.0506 \\
    Student & 0.3503 & 0.1078 & 0.1265 & 0.0416 \\
    \midrule
    \textit{Improv.T} & 8.85\% & 14.54\% & 10.76\% & 13.88\% \\
    \textit{Improv.B} & 8.59\% & 7.22\% & 4.42\% & 7.71\% \\
    \bottomrule
  \end{tabular}
\end{table}

All results are the average of five iterations and statistically significant with p=0.01.
The proposed approach (BD) still outperforms the best competitor (CD) both on ML100K and AMusic.
Moreover, the ranking performance of the teacher also increases with BD.

\vspace{20cm}